\documentstyle[12pt,aaspp4]{article}

\begin{document}

\title{900-1200\AA~ interstellar extinction in the Galaxy, LMC, and SMC
\footnote{Based on observations made with the
NASA-CNES-CSA Far Ultraviolet Spectroscopic Explorer. FUSE is operated for
NASA by the Johns Hopkins University under NASA contract NAS5-3298}
}

\author{J.B. Hutchings, J. Giasson}
 \affil{Herzberg Institute of Astrophysics,
National Research Council of Canada,\\ Victoria, B.C. V8X 4M6, Canada\\
john.hutchings@nrc.ca}

\begin{abstract}

 We have derived far-ultraviolet extinction curves for stars in the
Galaxy, LMC, and SMC, from spectra of pairs of stars observed with 
the FUSE satellite, matched in spectral type and with different amounts 
of reddening. The stars were observed
for other FUSE programs and were not selected for large reddening.
Thus, uncertainties in the derived extinction curves depend 
on the photometric data. In the case of strongly reddened stars from
the Galaxy, the strong H$_2$ absorption at the shortest wavelengths is an
additional complication. These initial results indicate that there is a family
of extinction curves that rise strongly to the Lyman limit, and extend the
related extinction curves from longer wavelengths.

\end{abstract}

\it Subject headings: \rm dust,extinction --- Galaxies: ISM ---
Galaxies: local group --- galaxies: Magellanic Clouds --- Ultraviolet: ISM

\section{Introduction}

 The interstellar extinction down to the Lyman limit is of interest
for several reasons. Shortward of 1200\AA, the extinction rises strongly and
differences between galaxy environments appear to increase. This means that 
corrections to observed (UV restframe) fluxes are potentially very 
sensitive to the amount
and type of interstellar extinction. This is important in understanding the
energy budget of galaxies with young stellar populations, and at high
redshifts in identifying the full population of young galaxies.

  There have been many studies in the IUE/HST range (1200 - 3200\AA) where
it is known that the extinction curve varies widely between different
environments (see e.g. Cardelli, Clayton, and Mathis 1989: CCM; Gordon and
Clayton 1998). In particular, the UV extinction laws
differ stongly between the Galaxy, LMC, and SMC, in the sense that
lower abundance galaxies have higher UV extinction for a given visible
reddening, and less pronounced 2175\AA~ `bump'. This general rule is thought
to indicate that the ISM composition or range of grain sizes changes with
continued stellar enrichment. There are, however, variations within
the local galaxies that must reflect a range of local ISM conditions or
history. It is more difficult to obtain reddening for the much fainter
stars in the M31 group galaxies, but work by Bianchi et al (1996)
shows that the variations are seen there too.

  In the range below 1200\AA, observations are rare and the rising extinction
makes it difficult to get data on significantly reddened stars. The HUT
spectrograph on ASTRO yielded extinction curves in this range, 
but for only 2 stars (Clayton et al 1996). 
FUSE is more powerful, has higher resolution, and
is building up an archive of spectra suitable for this work. While there
are programs planned to observe reddened stars specifically for extinction
studies, we present in this paper an initial, but significant derivation 
of extinction from lightly reddened stars observed for other purposes. 
 This study describes some of the difficulties in deriving extinction
curves in this wavelength range, and presents some results from
the Galaxy and Magellanic Clouds, where the expected extinction
differences are seen and measured.

  The sample of stars was chosen to contain pairs matched in spectral type
and luminosity with different B-V, assumed to indicate a different
amount of extinction. Table 1 shows the stars selected from the FUSE PI
team database of OB stars taken for stellar wind and O VI studies. These
stars were selected for their spectral class and UV flux, and thus are
not highly reddened, except for the Galactic stars, where brighter stars
are available. All observations were taken with the large FUSE aperture
and the data were processed uniformly with the CALFUSE data pipeline.
The data were not obtained through the FUSE archive as most were still
proprietary, so the data used may possibly differ from that which were 
delivered to the archive.

   Since the Magellanic Cloud stars are lightly reddened, the precision
of the colour photometry is an important part of the error budget of this work.
We have used what we regard as the most reliable values from the literature
in Table 1, but in all cases we investigated the effects of photometric errors
within the range (typically a few hundredths of a magnitude) of values
published for the individual stars, or other stars from the same papers.
We also rely on spectral types to pick good pairs of stars to compare for
differential reddening, and a similar inspection of published spectral types
was made. The adopted spectral types are given in Table 1. As the FUSE archive
grows, more and better pairs will become available, so this list represents
only those that were suitable after the first year of operations.

\section{Data processing}

 The FUSE spectra were edited of airglow emission lines, where significant
(see Feldman et al 2001),
before further processing. Using model spectra of H$_2$ absorption
(Tumlinson, private communication), the
strongest features were edited out by placing a continuum across them.
These absorptions are signifcant at the shorter wavelengths (below 1100\AA),
and usually in stars with E$_{B-V}$=0.2 or more. See Hutchings et al
(2001) for an illustration of the H$_2$ spectrum. Strong wind lines, present 
in high luminosity O stars, were identified from the atlas of OB FUSE spectra
which the FUSE team is preparing (or as illustrated by Fullerton et al 2000),
and also edited out.

 The edited spectra were then ratioed with the less reddened star as
numerator, and the result was heavily smoothed (5\AA~ or more boxcar). This 
was done separately for all channels of FUSE data, so that each pair of 
stars yielded a set of 8 ratioed spectra. These were then plotted in 
inverse wavelength, normalised to unit E$_{B-V}$, as is conventional for 
extinction plots.

  This approach usually resulted in 8 non-continuous lines, although
all showed the expected rise with decreasing wavelength. The discontinuities
arise from the H$_2$ blanketing, which changes strongly with wavelength,
flux errors at the edges of channels, FUSE detector `worm' effect, and 
possibly extraction incompleteness. The small differences of reddening 
between the stars in our pairs
means that these are amplified by large factors (typically 30) in normalising
to unit E$_{B-V}$. Nevertheless, the average of each emsemble of 8 plots
appeared to represent credible extinction curves for the FUSE range.

  In order to treat each spectrum more uniformly across the whole FUSE range,
the spectra were edited of airglow and then all channels resampled to 1A
bins and overplotted. H$_2$ models were generated with the same sampling
and plotted for a range of absorbing columns. Some stars have had detailed
H$_2$ modelling by Tumlinson et al (2001). This process allowed us to
a) identify a smooth mean observed spectrum over the whole FUSE range, and
b) estimate the continuum level of the star by correlating the strengths
of the main H$_2$ absorption features, and interpolating across any strong wind lines.

 In this way, each spectrum was modelled as a continuum defined by 12 
points across 
the full FUSE wavelength range, and these were then divided and normalised.
The results were generally consistent with the separate channel plots,
but without the discontinuities.

 The absolute level of the extinction was derived by correcting the
numerator spectrum by a factor which amounts to the ratio of unreddened
V-band flux, taking into account the observed V magnitudes and B-V colours.
This is standard practice and treats the numerator star as having total
extinction equal to the difference of extinction between the two stars. 

  The results for the sample stars are shown in Figures 1 and 2. Because of 
the large multiplier applied for small extinction differences,
we repeated the processing for a set of V and B-V values for the stars
covering the range of published values, or their errors.
Figures 1 and 2 illustrate these, and show the
sensitivity of the derived extinction curves to these quantities.

    The figures plot the overall continuum 12-point `models'. The results
from the individual channel plots are illustrated for the nominal 
photometric values by the pairs of filled squares. These show the
end points of linear fits to the ensembles of 8 individual channel plots.
The agreement between
these and the plotted continuum models gives a good idea of the
reproducibility of our analysis by two different approaches.

\section{Discussion}

   The individual extinction curves in Figures 1-2 were assessed for 
their sensitivity to the uncertainties in continuum fitting, and
to the photometric values as described above. There is clearly a wide 
range of curves present, and
some are very dependent on the photometric precision. Figure 3 shows
our adopted mean curves for the three galaxies, derived from 
the individual ranges of extinction curves in Figures 1 and 2,
weighted by the colour difference for each pair. These mean curves
were calculated for 5 wavelength bins.

   The plots also show schematic longer wavelength extinction curves 
for the Galaxy, LMC, and SMC, that represent those found in the 
literature. In Figure 3 we indicate the full range of curves for 
the LMC and SMC from Misselt et al (1999) and Gordon and Clayton (1999), 
while for simplicity the smaller diagrams just show
`typical' curves for each galaxy.

  The curves in Figure 3 have `formal' errors per point which are 
about the separation between the three curves. However, the fact that the
curves are nearly parallel and do not cross, add significance to the
overall family of curves and their trend with host galaxy abundances.
Nevertheless, it is clear from Figures 1 and 2 that more reliable results
will require stars with more extinction, more detailed continuum modelling,
and more precise photometry.

   Clayton et al (1996) published extinction curves for two LMC stars
in this range from HUT spectra. These are more heavily reddened than ours,
and their extinction curves closely bracket our mean Galactic curve, while
lying above the CCM sketched model. However, the general steepening rise
is reproduced very closely.

   We thank Luciana Bianchi and Alex Fullerton for advice and assistance
with the data, spectral types, and H$_2$ modelling.

\clearpage
\centerline{References}

Bianchi L., Clayton G.C., Bohlin R.C., Hutchings J.B., Massey P.,
1996, ApJ, 471, 203

Cardelli J.A., Clayton G.C., and Mathis J.S., 1989, ApJ, 345, 245 (CCM)

Clayton G.C. et al, 1996, ApJ, 460, 313

Gordon K.D., and Clayton G.C., 1998, ApJ, 500, 816

Feldman P.D., Sahnow D.S., Kruk J.W., Murphy E.M., Moos H.W., 2001, Journal
of Geophysical Research 106, 8119

Fullerton A.W., et al 2000, ApJ, 538, L43

Hutchings J.B., Crampton D., Cowley A.P., Schmidtke P., Fullerton A.W., 2001,
AJ, in press

Misselt K.A., Clayton G.C., Gordon K.D., 1999, ApJ, 515, 128

Tumlinson J. et al, 2001, preprint

\clearpage

\centerline{\bf Captions to figures}

1. Extinction from Galactic star pairs as indicated. Schematic 
extinction curves for SMC, LMC, and CCM are shown in each panel
for comparison. The nominal values of $\Delta_{B-V}$ are labelled,
and the other values are given to show the sensitivity of the extinction
curve to the photometric precision. In the case of the B0.5 pair,
we also show the effect of a change of V magnitude with the same colour
difference. The black squares show the end values for the extinction
derived from the separate chennel plots (see text for description).
The agreement between the squares and the nominal colour difference plots
is an indication of the random errors in continuum fitting.

2. As for Figure 1, for star pairs in the LMC and SMC
   
3. Mean values from star pairs in the three host galaxies, from
Figures 1 and 2. The formal errors on the five values plotted for each
galaxy are $\pm$1.5 magnitudes, but the general behaviour of the
set of curves suggest that their shapes and separation are significant.
The star and open square symbols outline the extinction curves 
from 2 LMC stars by Clayton et al (1996). The longer wavelength 
schematic plots show the range of extinction curves derived in the SMC
and LMC from references given, and the solid line is the CCM curve for
Galactic extinction. Note that our results appear to extend the extinction
curves for the three galaxies in a straightofrward way. The structure
in the FUSE range curves is unlikely to be more than noise.

\clearpage

\scriptsize
\begin{deluxetable}{lllrlll}
\tablecaption{Sample stars}
\tablehead{\colhead{Name} &\colhead{Sp Type} &\colhead{M$_V$} &\colhead{B-V}
&\colhead{E$_{B-V}$} &\colhead{Max flux}\tablenotemark{1} &\colhead{Dataset}}
\startdata

{\bf GALAXY}\nl
\nl

HD15558 & O5IIIf &7.95 &0.4 &0.73 &9.19E-13 &P1170101\nl 
HD93843 &O5IIIfvar &7.33 &-0.03 &0.30 &8.36E-11 &P1024001\nl
\nl
HD99857 & B0.5Ib & 7.47 &0.16 &0.48 &1.36E-11 &P1024501\nl 
HD100276 &B0.5Ib & 7.22 & 0.041 &0.36 &2.62E-11 &P1024801\nl
\nl
HD148422 &B1Ia  & 8.64 &0.11 &0.38 &2.03E-12 &P1015001  \nl
HD163522 &B1Ia  &8.42 &-0.003 &0.27 &3.39E-12 &P1015801\nl
\nl
HD74711 & B1III & 7.119 &0.08 &0.35 &1.57E-11 &P1022501 \nl 
HD119069 &B1III  &8.41 & -0.2 &0.07 &6.23E-11 &P1014001\nl
\nl
\nl
{\bf LMC}\nl
\nl
SK-70D60 & O4V & 13.85 &-0.19 &0.145 &4.85E-13 &P1172001 \nl
SK-70D69 & O4V & 13.941 &-0.36 &0.025 &5.61E-13 &P1172101\nl
\nl
SK-68D52 &B0Ia &11.716 &-0.065 &0.235 &7.52E-13 &P1174001 \nl 
SK-67D76 &B0Ia & 12.42 & -0.13 &0.17 &7.00E-13 &P1031201\nl
\nl
SK-68D171 &B1Ia &12.03 &-0.08 &0.19 &6.86E-13 &A0490801 \nl
SK-67D14 &B1Ia &11.5 &-0.115 &0.155 &9.87E-13 &P1174201\nl
\nl
{\bf SMC}\nl
\nl
AV26  & O7III &12.55 &-0.19 &0.13 &5.84E-13 & P1176001 \nl
AV69  & O7III & 13.35 &-0.22 &0.10 & 5.34E-13 &P1150303\nl
\nl
AV238  &O9III &13.77 &-0.22 &0.09 &4.77E-13 &P1176601 \nl
AV378  &O9III & 13.88 &-0.24 &0.07 &5.38E-13 &P1150707\nl
\nl
AV372  &O9.5I &12.63 &-0.18 &0.13 &7.56E-13 &P1176501 \nl
AV321  & O9I  &13.88 &-0.21 &0.10 & 4.55E-13 &P1150606 \nl
\enddata
\tablenotetext{1}{Value of continuum maximum in FUSE range in
erg.cm$^{-2}$.\AA$^{-1}$.sec$^{-1}$}
\end{deluxetable}

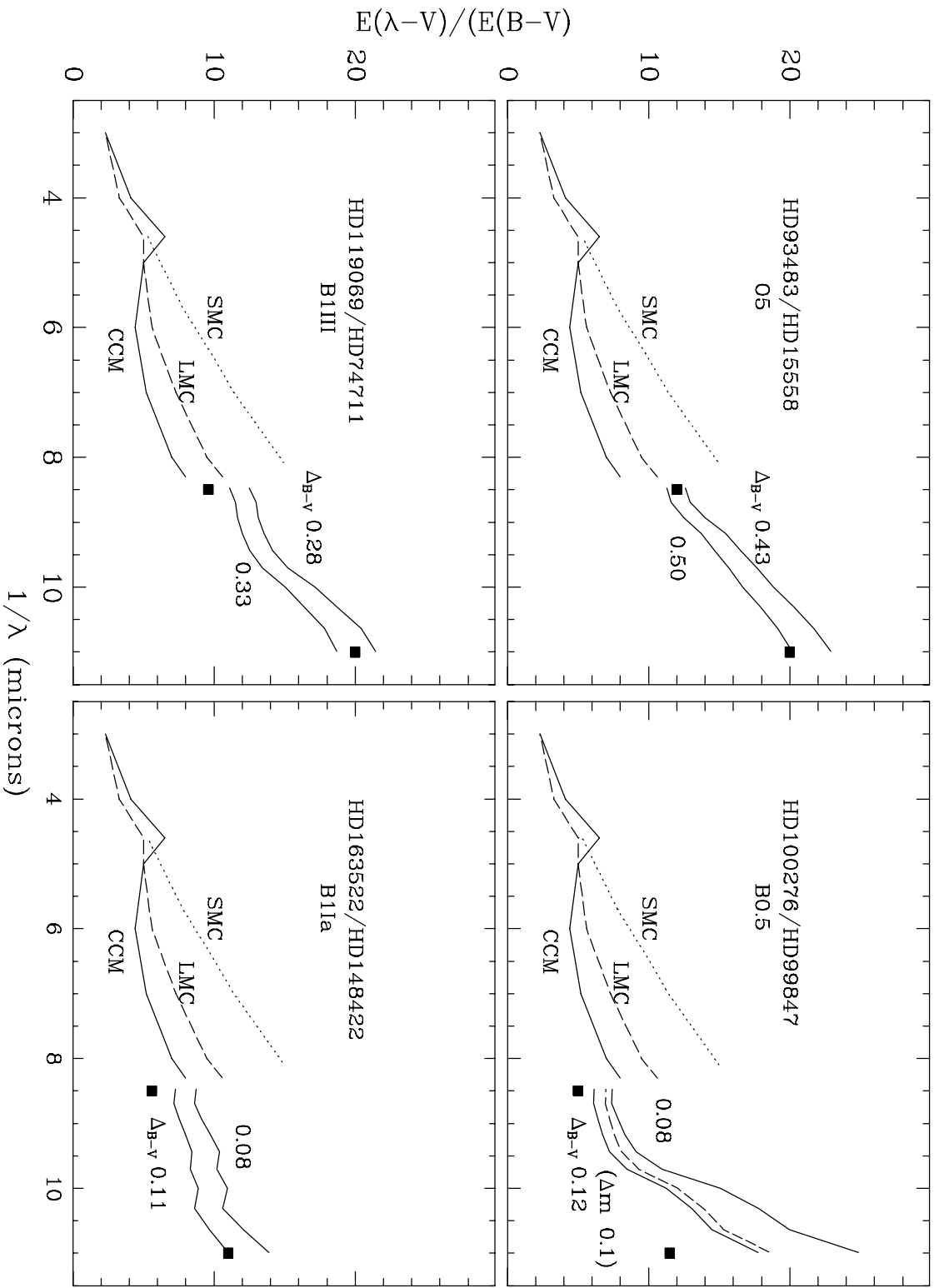
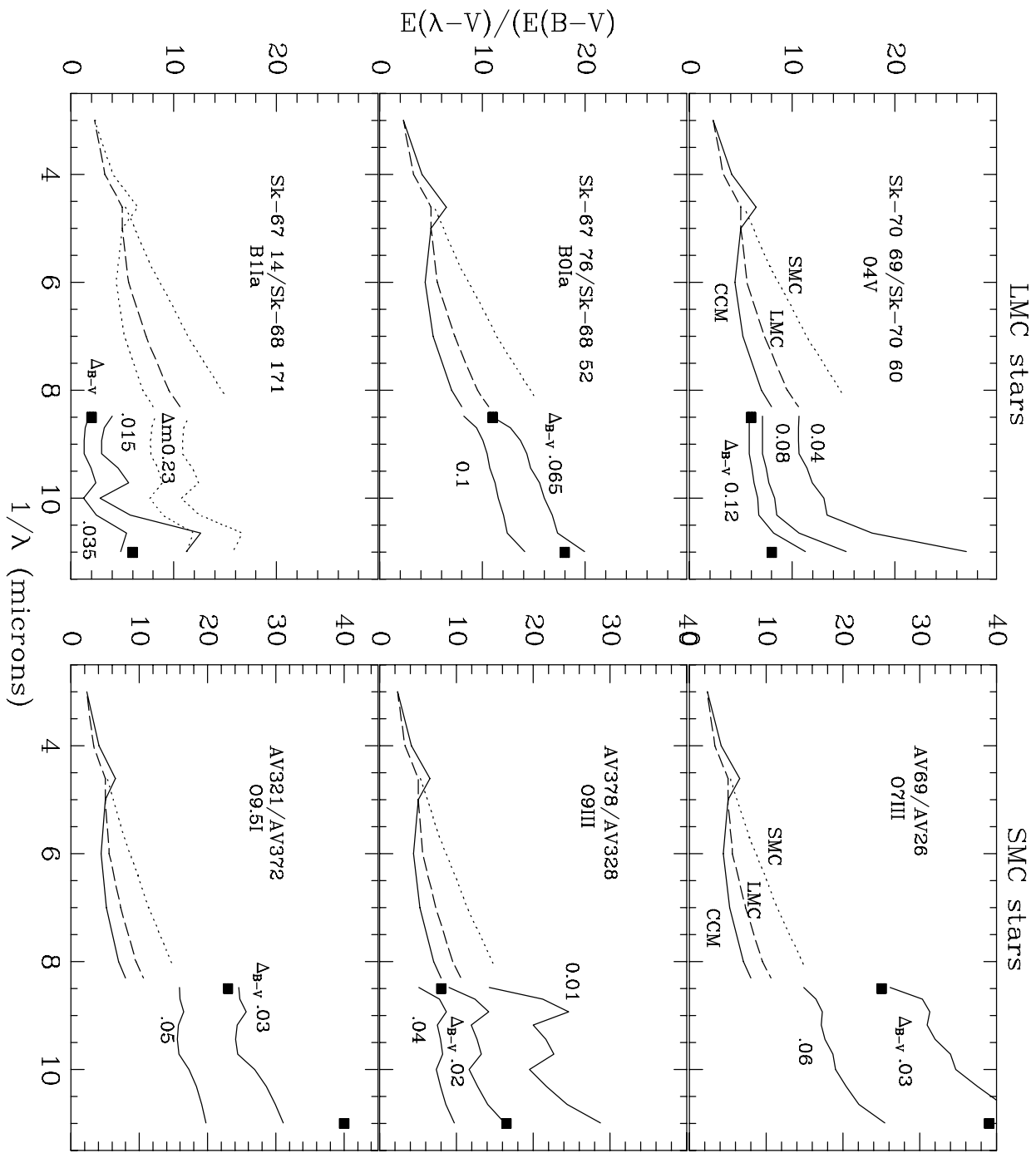
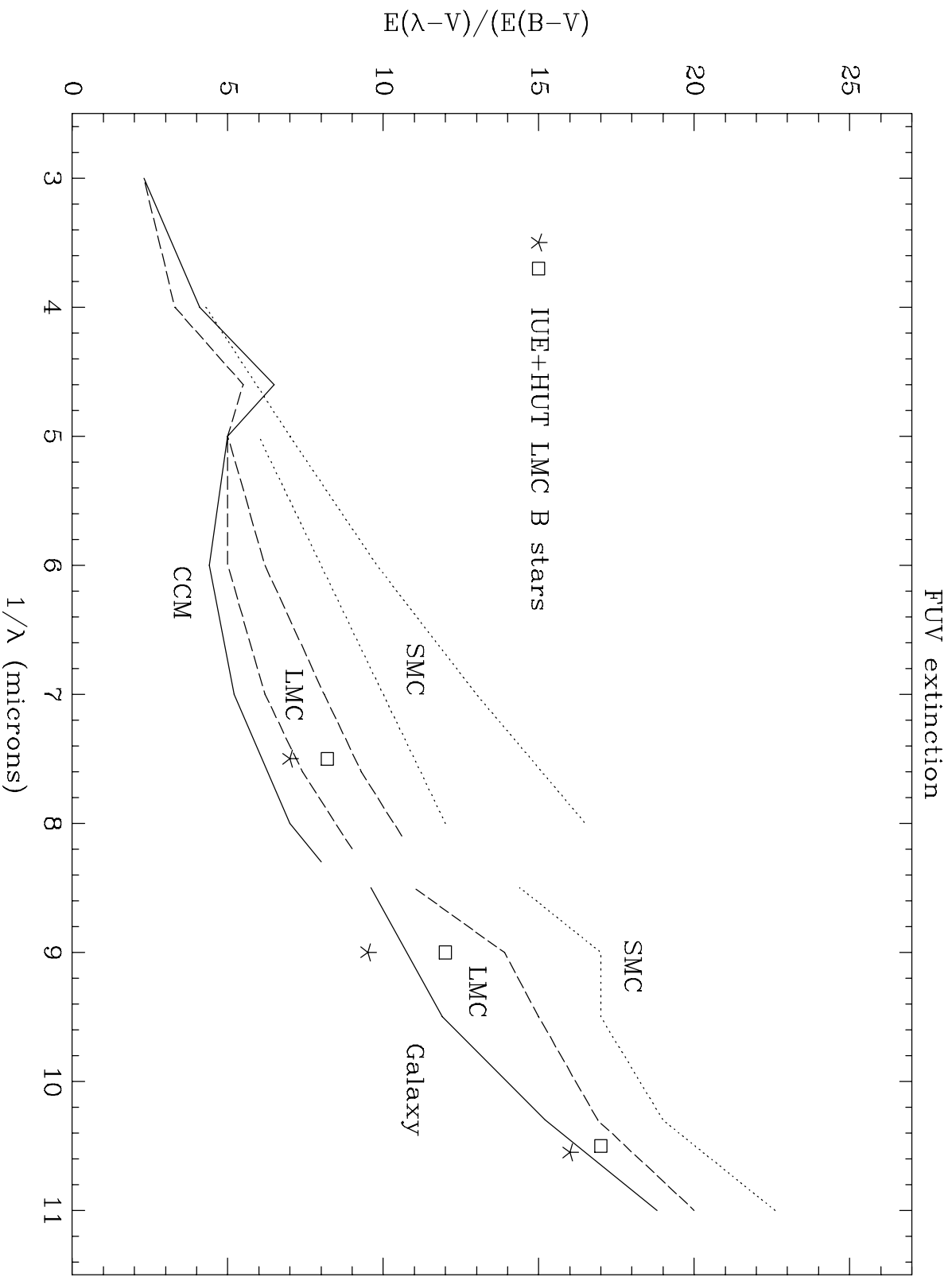
\end{document}